\documentclass[12pt,a4paper]{article}
\usepackage[T1]{fontenc}
\usepackage[latin1]{inputenc}
\usepackage{graphics}

\makeatletter

\providecommand{\LyX}{L\kern-.1667em\lower.25em\hbox{Y}\kern-.125emX\@}

\usepackage{graphics}

\newcommand{\be}{\begin{equation}}
\newcommand{\ee}{\end{equation}}
\newcommand{\bea}{\begin{eqnarray}}
\newcommand{\eea}{\end{eqnarray}}
\newcommand{\ra}{\rightarrow}

\newcommand{\half}{\raisebox{1pt}{$\scriptstyle \frac{1}{2}$}}
\newcommand{\quarter}{\raisebox{1pt}{$\scriptstyle \frac{1}{4}$}}

\newlength{\updownindent}
\newlength{\leftrightindent}
\makeatother

\begin{document}

\thispagestyle{empty} \renewcommand{\thefootnote}{\alph{footnote}}

{\par\raggedleft \texttt{MC-TH-01/07}~\\
\texttt{SHEP-01/17}~\\
\texttt{}~\\
\texttt{July 2001}\par}

\vspace{3cm}
{\par\centering \textbf{\large Higgs mass bounds in a Triplet Model}\\
 \vspace{0.2in} \textbf{J.R. Forshaw\( ^{1} \), D.A. Ross\( ^{2} \) and B.E.
White\( ^{1} \)}\\
 \vspace{0.2in} \( ^{1} \)\textit{Department of Physics and Astronomy,}\\
 \textit{University of Manchester,}\\
 \textit{Oxford Road,}\\
 \textit{Manchester M13 9PL, U.K.}\\
 \vspace{0.1in} \( ^{2} \)\textit{Department of Physics and Astronomy,}\\
 \textit{University of Southampton,}\\
 \textit{Southampton SO17 1BJ, U.K.}\\
 \vspace{0.15in}\par}
\bigskip{}

\begin{abstract}
\noindent We perform a global fit to high energy precision electroweak data
in a Higgs model containing the usual isospin doublet plus a real isospin triplet.
The analysis is performed in terms of the oblique parameters \( S \), \( T \)
and \( U \) and we show that the mass of the lightest Higgs boson can be as
large as 2 TeV.\newpage

\end{abstract}
\setcounter{page}{1}

\renewcommand{\thefootnote}{\arabic{footnote}} \setcounter{footnote}{0}

\section{Introduction}

\label{sec_intro}

With the high-energy measurements of electroweak (EW) observables by LEP and
SLD \cite{LEP}, an impressive level of precision has been achieved, in many
cases to 0.1\%. These have confirmed the Glashow-Salam-Weinberg model of EW
broken gauge symmetry with great certainty. What remains is to discover the
nature of the symmetry breaking. If it is the Standard Model (SM) Higgs, i.e.
a complex isospin doublet, the hard empirical lower bound we have on its mass
is the current 113 GeV from the LEP search \cite{LEP}.

The other empirical bound on the SM Higgs is less direct. It involves constraining
the radiative corrections to EW observables already measured. One feature of
the virtual Higgs corrections is that, to a very good approximation, they are
\textit{oblique}, i.e. they appear only in the corrections to propagators of
the EW gauge bosons \cite{Lynn}. It happens that the important effects can
be summarised by two parameters, \( S \) and \( T \) \cite{PT}, which on
the one hand can be calculated in the SM, and on the other hand fitted globally
to all current precision data. Another important feature is that the dependence
of \( S \) and \( T \) on the SM Higgs mass \( m_{h} \) is logarithmic. However,
although the \( m_{h} \)-dependence is weak, one can nevertheless put an upper
limit \( m_{h}<165 \) GeV at the 95\% confidence level \cite{LEP}.

Such an upper limit on \( m_{h} \) is necessarily model-dependent, in the sense
that it applies only to the minimal SM scenario. In this letter, we consider
a Higgs model \cite{RV,BH} (TM) with the SM complex doublet plus a real triplet
of scalars. The physical spectrum contains two extra states, namely another
neutral \( k^{0} \) and a charged \( h^{\pm } \). The model violates the custodial
symmetry responsible for the tree-level relation \be
  \rho \; \equiv \; \frac{m_{W}^{2}}{m_{Z}^{2} c_{W}^{2}} \; = \; 1, \ee
where \( c_{W}=\cos \theta _{W}=g/\sqrt{(g^{2}+g^{\prime 2})} \) but by making
the triplet vev small the relation can be satisfied to within the experimental
uncertainties. 

In this small vev approximation, we shall show that the tree level corrections
can be absorbed into a shift in \( T \) which is of the correct sign to partially
cancel the SM Higgs contribution. We also compute the one-loop corrections where
the two new particles contribute directly to the oblique parameters: \( S \)
and \( U \) are largely unaffected whilst \( T \) receives a correction depending
on the ratio \( (\Delta m/m_{Z})^{2} \), where \( \Delta m \) is the mass
splitting between the \( k^{0} \) and \( h^{\pm } \). Like the tree-level
correction, this is also of the correct sign to partially cancel the SM Higgs
contribution. We demonstrate that it is consequently possible to relax the upper
limit on \( m_{h} \) to values as large as 2 TeV.

The discussion below is arranged as follows: Section~\ref{sec_model} gives
the Lagrangian for the triplet model~(TM) and a brief description of its spectrum;
in Section~\ref{sec_oblique} we define the parameters \( S \), \( T \) and
\( U \) then show how any given EW observable depends on them, and give the
result of their calculation in the SM and the TM; in Section~\ref{sec_zfitter}
we compare the SM and TM calculations with our fit to the EW data. Finally,
we make some concluding remarks.

\section{The Model}

\label{sec_model}

The Lagrangian for the model, containing one complex Higgs doublet and one real
triplet is \bea
  \mathcal{L}(\Phi, H)  &=&  (D_\mu {}\Phi)^{\dagger }(D^{\mu }\Phi) \; + \; \half (D_\mu {}H)^{\dagger }(D^{\mu }H)
\; - \; V(\Phi, H), \nonumber \\
 V(\Phi, H) & =& \mu_1^{2} \, \Phi^{\dagger }\Phi \; + \; \half \mu_2^{2} \,
H^{\dagger }H \nonumber \\
 & +& \lambda_1 \, (\Phi^{\dagger }\Phi)^{2} \; + \; \quarter \lambda_2 \, (H^{\dagger }H)^{2} \nonumber \\
 &+& \half \lambda_3 \, (\Phi^{\dagger }\Phi)(H^{\dagger }H) \; + \; \lambda_4
\, v \, H_U^{i} \, \Phi^{\dagger }\sigma^{i} \Phi. \eea

The field components are, including the neutral components' vevs, \be
  \Phi \; = \; \left(\begin{array}{c}\phi^{+} \\
   \frac{1}{\sqrt{2}}(v\, + \, \phi^{0}_{R} \, + \, i \phi^{0}_{I})
   \end{array}\right), \phantom{******} H \; = \; \left(\begin{array}{c}\eta^{+} \\
   \half v t_{\beta
   }\, + \, \eta^{0} \\ -\eta^{-}\end{array}\right). \label{field_compts}\ee

In the above, \( \sigma ^{i} \) are the Pauli matrices and \( t_{\beta }=\tan \beta  \).
\( H \) is even under charge conjugation, i.e. \be
  H \; = \; H_c \; = \; C \, H^{\ast}, \ee
where \be 
   C \; = \; \left(\begin{array}{rrr}0 & 0 & -1 \\
   0 & 1 & 0 \\ -1 & 0 & 0\end{array}\right). \ee
and can be cast into a form, \( H_{U} \), involving only real fields, by the
unitary transformation \be
  H_U \; = \; U^{\dagger }H. \ee
where \be
 U \; = \; \frac{1}{\sqrt{2}} \left(\begin{array}{rrr} 1 & -i & 0
   \\ 0 & 0 & \sqrt{2} \\
   -1 & -i & 0 \end{array}\right). \ee

Expanding about the vacuum by substituting eq.~(\ref{field_compts}) into the
Lagrangian, we can analyse the mass spectrum. One finds two charged Higgs states.
The first, \( g^{\pm } \), is massless and is the Goldstone to be eaten by
the \( W^{\pm } \). The second we call \( h^{\pm } \), having mass \( m_{c} \)
such that \be
m_c^{2} \; = \; \frac{\lambda_{4} v^{2}}{s_{\beta} c_{\beta}} \sim \frac{\lambda_{4} v^{2}}{\beta}.
\ee
In terms of the original doublet and triplet charged components these are \be
  \left(\begin{array}{c} g^{\pm} \\ h^{\pm} \end{array}\right) \; = \; \left(\begin{array}{rr} c_{\beta }& s_{\beta }\\ -s_{\beta }& c_{\beta }\end{array} \right)
\left(\begin{array}{c} \phi^{\pm} \\ \eta^{\pm} \end{array}\right). \ee
In the charge neutral sector we have a CP-odd massless state which is the Goldstone
\( g^{0} \) to be eaten by the \( Z^{0} \):\be
  g^{0} \; = \; \phi^{0}_I. \ee
Finally, there are two CP-even states, called \( h^{0} \) and \( k^{0} \),
having mass \( m_{h} \) and \( m_{k} \) respectively. In terms of the original
doublet and triplet components there is generically a mixing: \be
  \left(\begin{array}{c} h^{0} \\ k^{0} \end{array}\right) \; = \; \left(\begin{array}{rr}
   c_{\gamma }& s_{\gamma }\\ -s_{\gamma }& c_{\gamma }\end{array} \right) \left(\begin{array}{c} \phi^{0} \\ \eta^{0} \end{array}\right).
\ee
For simplicity, we shall only consider the case of zero mixing, \( \gamma =0 \),
leading to masses \bea
  m_h^{2} &=&  2 \, \lambda_1 \, v^{2}, \\
 m_k^{2} &=& \half \, \lambda_2 \, (v t_\beta{})^{2} \; + \; \frac{\lambda_{4} v^{2}}{t_{\beta}}.
\eea

The model has six parameters, \( \mu _{1,2} \) and \( \lambda _{1,2,3,4} \),
or alternatively \( v \), \( \beta  \), \( m_{c} \), \( m_{h} \), \( m_{k} \)
and \( \gamma  \). However, since we assume zero mixing between the neutral
CP even Higgses this reduces to five parameters because \( \lambda _{3}=2\lambda _{4}/t_{\beta } \).

The most important tree-level predictions of the model are the masses for the
\( W^{\pm } \) and \( Z^{0} \), which are \be
  m_W \; = \; \frac{g v}{2 c_{\beta}}, \phantom{******} m_Z \; = \; \frac{g v}{2 c_{W}},
\label{tree_mass}\ee
where \( c_{\beta }=\cos \beta  \). The expression for \( m_{Z} \) is identical
to the SM with just the doublet, while \( m_{W} \) is increased relative to
the SM. This gives a tree-level \( \rho  \)-parameter: \be
  \rho \; \equiv \; \frac{m^{2}_{W}}{m^{2}_{Z} c^{2}_{W}} \; = \; \frac{1}{c^{2}_{\beta}}. \label{newrho} \ee
Thus \( s_{\beta } \) has to be less than a few percent in order to have a
realistic phenomenology.

An important issue is that taking \( s_{\beta }\ra 0 \), keeping all other
parameters in the Lagrangian fixed, is a decoupling limit such that all effects
of the extra Higgs triplet on EW observables become negligible. However, and
as we shall soon show, it is quite possible to have interesting phenomenology
with small but non-zero \( \beta  \).

\section{Oblique Corrections}

\label{sec_oblique}

The parameters \( S \), \( T \) and \( U \) are defined as \bea
  \alpha \, S  &=&  \frac{4 s^{2}_{W} c^{2}_{W}}{m^{2}_{Z}} \; \left( \Delta \Pi^{ZZ}(m_Z)
\; - \; \frac{c^{2}_{W} - s^{2}_{W}}{s_{W} c_{W}} \; \Delta \Pi^{\gamma Z}(m_Z)
\; - \; \Delta \Pi^{\gamma\gamma}(m_Z) \right), \nonumber \\
 \alpha \, T  &=& \frac{1}{m^{2}_{W}} \; \left( \Pi^{WW}(0) \; - \; c^{2}_W \; \Pi^{ZZ}(0)
\right), \nonumber \\ 
\alpha \, (S+U) &=& 4 s_W^2 \left( \frac{\Delta \Pi^{WW}(m_W)}{m_W^2}-\frac{c_W}{s_W} \frac{\Delta \Pi^{\gamma Z}(m_Z)}{m_Z^2} - \frac{\Delta \Pi^{\gamma \gamma}(m_Z)}{m_Z^2} \right),
\label{sandt}\eea
where \( \Delta \Pi (k)=\Pi (k)-\Pi (0) \). The functions \( \Pi (k) \) are
the coefficients of the metric in the one-loop gauge boson inverse propagators:
\be
  \Pi_{\mu\nu}(k) \; = \; g_{\mu\nu} \; \Pi(k). \ee

Predictions for EW observables, which we write generically as \( \mathcal{O} \),
can be written in terms of \( S \), \( T \) and \( U \). If we are just considering
the prediction of the SM, we can write \bea
  \mathcal{O}_{SM}(m_h)  &=& \mathcal{O}_{SM}(m_h^{ref}) \nonumber \\
 &+& \alpha \, A_{SM} \, \Delta S_{SM}(m_h,m_h^{ref}) \nonumber \\
 &+& \alpha \, B_{SM}\, \Delta T_{SM}(m_h,m_h^{ref}) \nonumber \\
&+& \alpha \, C_{SM} \, \Delta U_{SM}(m_h,m_h^{ref}). \eea
Here, \( \mathcal{O}_{SM} \) is the one-loop SM prediction for the observable,
in terms of the input parameters \( \alpha (0) \), \( m_{Z} \), \( G_{\mu } \),
\( m_{t} \), \( \alpha _{s}(m_{Z}) \) and \( \Delta \alpha _{had}^{(5)}(m_{Z}) \).
The first term on the r.h.s. is the SM prediction evaluated at a fixed reference
Higgs mass, which is arbitrary. The coefficients \( A_{SM} \), \( B_{SM} \)
and \( C_{SM} \) are process dependent but independent of the new physics (which
in this case is that of the Higgs). \( \Delta S_{SM} \), \( \Delta T_{SM} \)
and \( \Delta U_{SM} \) are the contributions to \( S \), \( T \) and \( U \)
after subtracting their value at the reference Higgs mass, i.e. they are defined
to vanish when \( m_{h}=m_{h}^{ref} \). In this way one can quantify the effect
of varying the Higgs mass on the observable simply in terms of \( \Delta S_{SM} \),
\( \Delta T_{SM} \) and \( \Delta U_{SM} \).

If now we consider the prediction of the TM, some modifications are required.
In this case a general observable (setting \( \gamma =0 \)) is written as follows:
\bea
  \mathcal{O}_{TM}(m_h, m_k, m_c, \beta) &=& \mathcal{O}_{SM}(m_h^{ref}) \nonumber \\
 &+& A_{SM} \, ( \alpha \; \Delta S_{SM}(m_h,m_h^{ref}) \; + \; \alpha \; S_{TM}(m_k, m_c) )
\nonumber \\
&+& B_{SM} \, ( \alpha \; \Delta T_{SM}(m_h,m_h^{ref}) \; + \; \alpha \; T_{TM}(m_k, m_c) + \delta_{tree}(\beta) ),
\nonumber \\ &+& C_{SM} \,( \alpha \; \Delta U_{SM}(m_h,m_h^{ref})\; + \; \alpha \; U_{TM}(m_k,m_c)).\label{tm_oblique}\eea
Here we have extra contributions, denoted with $TM$ in subscript, coming from
the extra \( k^{0} \) and \( h^{\pm } \) loops. Since we are taking \( \beta ^{2} \)
to be small, there are some simplifications. To the accuracy we require, \( O(\alpha \beta ^{2}) \)
may be neglected, as may \( O(\beta ^{4}) \). Therefore, we may evaluate our
one-loop corrections, themselves of \( O(\alpha ) \), at \( \beta =0 \), so
that the coefficients of \( S_{TM} \) , \( T_{TM} \) and \( U_{TM} \) are
the same as in the SM. The only \( \beta  \)-dependence takes the form of \( O(\beta ^{2}) \)
corrections that occur at tree-level and these are contained in the correction
\( \delta _{tree}(\beta ) \). There appear to be two distinct types of contribution
to \( \delta _{tree}(\beta ) \):

\begin{enumerate}
\item Direct tree-level corrections. In our case, only one observable, \( m_{W} \),
has a direct tree-level correction, as seen in eq.(\ref{newrho}). This is because
it is the only high-energy EW observable we shall fit to which involves the
\( W \) boson at tree level. 
\item Indirect tree-level corrections: All the EW observables (except \( m_{W} \)
as we have just mentioned) can be written at tree-level in terms of \( \alpha  \),
\( m_{Z} \) and \( s_{W} \), none of which depends directly on \( \beta  \)
in the TM. However, \( s_{W} \) is constrained using the input datum \( G_{\mu } \)
which does itself have a dependence on \( \beta  \). At tree level, this is
\be
    \sqrt{2} \, G_\mu {}\; = \; \frac{g^{2}}{4 m^{2}_{W}} \; = \; \frac{4 \pi \alpha}{m^{2}_{Z}} \, \frac{c^{2}_{\beta}}{\sin^{2}2\theta_{W}}. \ee
\end{enumerate}
All observables we consider receive an indirect shift whilst only the \( W \)
mass picks up a direct shift. However, since the shift is essentially oblique
it can in all cases be absorbed into a shift in \( T \), as we anticipated
in eq.(\ref{tm_oblique}). In particular it is straightforward to show that,
to leading order, \begin{equation}
\label{tree}
\delta _{tree}=\beta ^{2}.
\end{equation}
 For use later we define \begin{eqnarray}
S &  & =\Delta S_{SM}+S_{TM}\nonumber \\
\alpha \: T &  & =\alpha \: \Delta T_{SM}+\alpha \: T_{TM}+\delta _{tree}\label{SandT} \\
U &  & =\Delta U_{SM}+U_{TM}\nonumber 
\end{eqnarray}

Calculation of the SM Higgs boson contributions to \( S \) and \( T \) are
as follows (evaluated at \( m_{Z} \)): \begin{eqnarray} S_{\rm SM} & = & \frac{1}{\pi} \left[ \frac{3}{8}\,\frac{m^2_h}{m^2_Z} - \frac{1}{12} \, \frac{m^4_h}{m^4_Z}~ \right. \nonumber \\ & + & \frac{m^2_h}{m^2_Z} \, \log\left(\frac{m^2_h}{m^2_Z}\right) \left( \frac{3 m^2_Z - m^2_h}{4 m^2_Z} \; + \; \frac{1}{24} \, \frac{m^4_h}{m^4_Z} \; + \; \frac{3 m^2_Z}{4(m^2_Z - m^2_h)} \right) \nonumber \\ & + & \left(1 \; - \; \frac{1}{3} \, \frac{m^2_h}{m^2_Z} \; + \; \frac{1}{12} \, \frac{m^4_h}{m^4_Z} \right) \frac{m_h}{m_Z^2} \nonumber \\ & \times & \left. \left\{ \begin{array}{l} \sqrt{4m^2_Z - m^2_h} \, \tan^{-1} \sqrt{\frac{4m^2_Z - m^2_h}{m^2_h}}\; ; \phantom{**!!}\; m_h < 2m_Z \\ \sqrt{m^2_h - 4m^2_Z} \, \log\left(\frac{2m_Z}{m_h + \sqrt{m_h^2 - 4m_Z^2}}\right); \; m_h > 2m_Z \end{array} \right\} \right], \\ T_{\rm SM} & = & \frac{3}{16 \pi} \, \frac{1}{s^2_W c^2_W} \left[ \frac{m^2_h}{m^2_Z - m^2_h} \, \log\left(\frac{m^2_h}{m^2_Z}\right) \; - \; \frac{c^2_W m^2_h}{c^2_W m^2_Z - m^2_h} \, \log\left(\frac{m^2_h}{c^2_W m^2_Z}\right) \right]. \end{eqnarray}
We do not show \( U_{SM} \) since is depends very weakly on \( m_{h} \).

The TM contributions, to leading order in \( \beta  \), are (see Appendix)
\bea
S_{TM}&=&  0, \\
T_{TM} &=& \frac{1}{8\pi} \, \frac{1}{s^{2}_{W} c^{2}_{W}} \left[ \frac{m^{2}_{k} + m^{2}_{c}}{m^{2}_{Z}} \;
- \; \frac{2 m^{2}_{c} m^{2}_{k}}{m^{2}_{Z}(m^{2}_{k} - m^{2}_{c})} \log\left(\frac{m^{2}_{k}}{m^{2}_{c}}\right)
\right], \nonumber \\
 &\simeq &  \frac{1}{6\pi} \, \frac{1}{s^{2}_{W} c^{2}_{W}} \; \frac{(\Delta m)^{2}}{m^{2}_{Z}}. \nonumber \\ U_{TM} &=& -\frac{1}{3 \pi} \left( m_k^4 \log \left( \frac{m_k^2}{m_c^2} \right) \frac{ (3 m_c^2-m_k^2)}{(m_k^2-m_c^2)^3} + \frac{5(m_k^4+m_c^4)-22 m_k^2 m_c^2}{6(m_k^2 - m_c^2)^2} \right) + O(m_Z/m_c) \nonumber \\ &\simeq & \frac{\Delta m}{3 \pi m_c}. \eea

Notice that the TM contribution to \( S \) is zero to this order. The TM contribution
to \( T \) is positive and, in the approximation of \( \Delta m=m_{k}-m_{c}\ll m_{c} \),
has the rough power dependence shown above. \( U \) also vanishes when \( \Delta m\to 0 \),
and falls to zero at large triplet masses. In particular, it has a negligible
effect on all the results we shall subsequently show provided \( m_{k},\, m_{c}>1 \)
TeV.

We have thus shown that the TM generates a positive correction to \( T \) due
both to tree-level mixing and quantum loops. As we shall demonstrate in the
next section, this allows us to compensate for an increase in the doublet Higgs
mass thus releasing the SM upper bound. 

We note that the quantum corrections are important for \( \Delta m\sim m_{Z} \)
and that this is possible provided \( \lambda _{2}\gg \lambda _{4} \), e.g.
\( \lambda _{4}\sim \beta  \), \( \lambda _{3}\sim 1 \), \( \lambda _{2}\sim 1/\beta ^{2} \)
is a scenario which would lead to triplet bosons of mass \( \sim v \). In such
cases, \( \lambda _{2} \) is large and the Higgs sector would become non-perturbative.
More naturally, the triplet Higgs bosons are of mass \( \sim v/\beta  \) and
the mass splitting is much less than \( m_{Z} \). In this case, the principal
contribution will arise from the tree-level mixing.

\section{Comparison with Data}

\begin{figure}
{\par\centering \resizebox*{8cm}{8cm}{\includegraphics{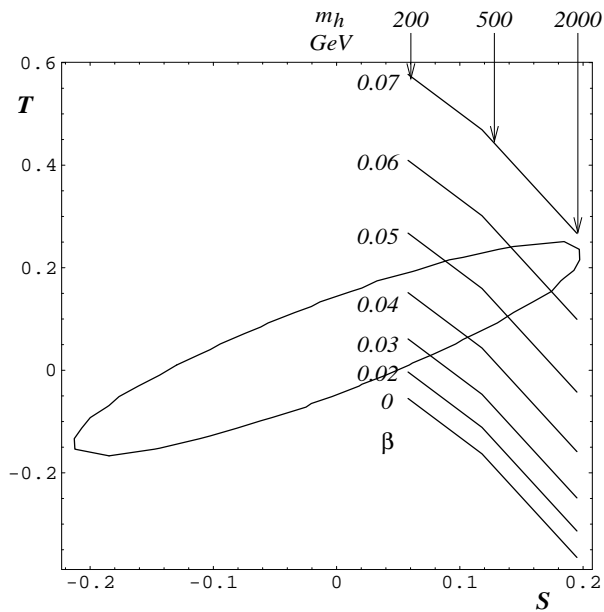}} \par}

\caption{Ellipse encloses the region allowed by data. Curves show results in the TM
for various values of \protect\( \beta \protect \) and various doublet Higgs
masses. \protect\( \Delta m=0\protect \) and \protect\( U=0\protect \) in
this plot.\label{Fig-beta}}
\end{figure}

\begin{figure}
{\par\centering \resizebox*{8cm}{8cm}{\includegraphics{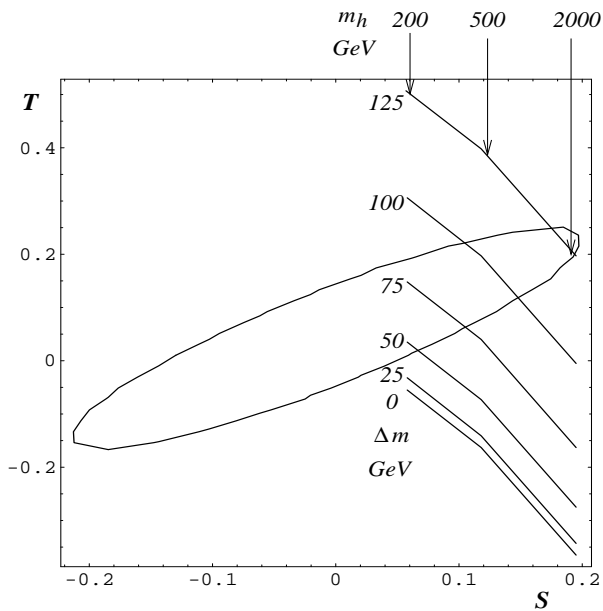}} \par}

\caption{Ellipse encloses the region allowed by data. Curves show results in the TM
for various mass splittings and various doublet Higgs masses. \protect\( \beta \protect \)
and \protect\( U\protect \) are assumed to be negligibly small in this plot.\label{Fig-dm}}
\end{figure}

\label{sec_zfitter}Using the program \texttt{ZFITTER} \cite{ZFITTER} \texttt{}we
compute a total of 13 standard observables\footnote{%
They are those listed in Table 41 of reference \cite{LEP}.
} with \( m_{h}^{ref}=100 \) GeV, \( m_{t}=174.3 \) GeV, \( G_{\mu }=1.6639\times 10^{-5} \)
GeV\( ^{2} \), \( m_{Z}=91.1875 \) GeV, \( \alpha _{s}= \)0.119 and \( \Delta \alpha _{had}^{(5)}(m_{Z}) \)=
0.02804. These results then determine the allowed region in \( S-T \) parameter
space. This is represented by the interior of the ellipse shown in Figures \ref{Fig-dm}
and \ref{Fig-beta}. The ellipse corresponds to a total chi-squared of 26.3
for the 17 measurements used. We have investigated variations in the ellipse
as the input parameters \( m_{t} \), \( \alpha _{s} \) and \( \Delta \alpha _{had}^{(5)} \)
are varied within their errors: a smaller value of \( \alpha _{s}=0.117 \)
is slightly favoured, varying \( m_{t} \) \( (\pm 5.1 \) GeV) leads to a shift
\( \pm 0.05 \) in \( T \), whilst varying \( \Delta \alpha _{had}^{(5)} \)
\( (\pm 0.00065 \)) leads to a shift of \( \pm 0.05 \) in \( S \).

In Figure \ref{Fig-beta} each line now shows the TM at a particular value of
\( \beta  \) for \( \Delta m=0 \) (which turns off the quantum corrections)
and \( m_{h} \) varying from 200 GeV to 2 TeV. We see that even in the absence
of quantum corrections the TM is able to accommodate any \( m_{h} \) up to
around 2 TeV and the mixing angle \( \beta  \) cannot be much larger than 0.07.

In Figure \ref{Fig-dm} each line shows the TM result as \( m_{h} \) is varied,
as before, at fixed \( \Delta m \). \( \beta  \) is assumed to be negligibly
small in this plot (which turns off the tree-level correction \( \delta _{tree} \))
and as a result the \( \Delta m=0 \) line is identical to that which would
arise in the SM. Clearly the quantum corrections contribute to \( T \) so as
to allow any \( m_{h} \) up to around 2 TeV and the mass splitting \( \Delta m \)
cannot be much larger than 125 GeV.

\section{Conclusions}

We have shown that it is quite natural in the triplet model for the lightest
Higgs boson to have mass as large as 1 TeV. Although quantum corrections could
play an important role in pushing up the Higgs mass we have shown that it is
perhaps most natural to do this through tree-level corrections which arise due
to mixing in the charged Higgs sector.

\section*{Appendix}

Here we give a few details on the calculation of \( S \), \( T \) and \( U \)
in the triplet model. Starting from their definitions in eq.~(\ref{sandt})
we can write them in terms of the standard functions, \( A \) and \( B_{22} \)
(up to order \( \beta ^{2} \) corrections): \bea
  S_{TM} &=&  0, \\
 T_{TM} &=& \frac{1}{4\pi} \, \frac{1}{s^{2}_{W} c^{2}_{W} m_Z^2} \left( 4 B_{22}(0; m_c, m_k)
\; - \; A(m_c) \; - \; A(m_k) \right), \nonumber \\ U_{TM} &=& \frac{4}{\pi} \left( \frac{B_{22}(m_W^2;m_c,m_k) - B_{22}(0;m_c,m_k)}{m_W^2}- \frac{B_{22}(m_Z^2;m_c,m_c) - B_{22}(0;m_c,m_c)}{m_Z^2} \right) \nonumber \eea
where \bea  \frac{i}{(4 \pi)^{2}} \; A(m) &=& \mu^{4-D} \int \frac{d^D k}{(2 \pi)^D} \frac{1}{k^{2} - m^{2}+i \epsilon}, \\ \frac{i}{(4 \pi)^{2}} \;
g^{\mu\nu} \; B_{22}(p^2; m_1, m_2) &=& \mu^{4-D} \int \frac{d^D k}{(2 \pi)^D}
\frac{k^{\mu} k^{\nu}}{(k^{2} - m_{1}^{2}+i \epsilon)((k-p)^{2} - m_{2}^{2}+i \epsilon)}.
\nonumber \\ \eea
These can be evaluated using dimensional regularisation \cite{integrals}, e.g.
\bea
A(m)  &=& m^{2} \, \left(\frac{1}{\epsilon} \; - \; \gamma_E \; + \; 1 \; - \; \log\left(\frac{m^{2}}{4\pi\mu^{2}}\right)
\right),\\
 B_{22}(0; m_1, m_2)  &=&  \frac{1}{4} \left[ \left(\frac{1}{\epsilon} \; -
\; \gamma_E \; + \; \frac{3}{2} \right) (m^{2}_1 \, + \, m^{2}_2) \right.\nonumber \\
 & -& \left. \frac{1}{m_{1}^{2} \, - m_{2}^{2}} \; \left( m_1^{4} \, \log\left(\frac{m_{1}^{2}}{4\pi\mu^{2}}\right) \;
- \; m_2^{4} \, \log\left(\frac{m_{2}^{2}}{4\pi\mu^{2}}\right) \right) \right].
\eea

\section*{Acknowledgements}

We should like to thank Nick Evans and Chris Llewellyn-Smith for helpful discussions.

\end{document}